\documentclass[twocolumn]{revtex4}
\usepackage[dvips]{graphicx}
\usepackage{dcolumn}
\usepackage{amsmath}
\usepackage{bm}
\begin{document}
\title{Development of displacement- and frequency-noise-free
interferometer in 3-D configuration for gravitational wave detection}

\author{Keiko Kokeyama}
\email{keiko.kokeyama@nao.ac.jp}
\affiliation{
Graduate School of Humanities and Sciences, 
Ochanomizu University, 2-1-1, Otsuka, Bunkyo-ku,
Tokyo 112-8610 Japan}

\author{ Shuichi Sato}
\affiliation{Faculty of Engineering, Hosei University,
3-7-2, Kajino-cho, Koganei, Tokyo 184-8584 Japan}

\author{Atsushi Nishizawa}
\affiliation{Graduate School of Human and Environmental Studies
Kyoto University, Kyoto 606-8501 Japan}

\author{Seiji Kawamura}
\affiliation{TAMA project, National Astronomical Observatory of Japan,
2-21-1, Mitaka, Osawa, Tokyo 181-8588 Japan}

\author{Yanbei Chen}
\affiliation{Theoretical Astrophysics, California Institute of Technology,
Pasadena, California 91125 USA}

\author{Akio Sugamoto}
\affiliation{Ochanomizu University, 2-1-1, Otsuka, Bunkyo-ku,
Tokyo 112-8610 Japan}

\begin{abstract}
The displacement- and frequency-noise-free interferometer (DFI) is a multiple laser interferometer array for gravitational wave detection free from both the displacement noise of optics and laser frequency noise. So far, partial experimental demonstrations of DFI have been done in 2-D table top experiments. In this paper, we report the complete demonstration of a 3-D DFI. The DFI consists of four Mach-Zehnder interferometers with four mirrors and two beamsplitters. The displacement noises both of mirrors and beamsplitters were suppressed by up to 40 dB. The non-vanishing DFI response to a gravitational wave was successfully confirmed using multiple electro-optic modulators and computing methods. 
\end{abstract}

\maketitle
%
Gravitational waves are perturbations of the space-time
curvature propagating across the universe at the speed of light.
They propagate in space-time,
extending a proper distance between test masses
in one transverse direction,
and shortening in the other, orthogonal direction.
A laser-interferometric gravitational wave detector
measures length differences
between the optical components (test masses)
caused by the gravitational wave effect.
Various noise sources that limit the sensitivity
of the detector
must be removed to achieve the best sensitivity level,
as the interferometers must detect the very tiny signals
of length change.
The sensitivities of the ground-based detectors
are affected by seismic noise,
thermal noise, radiation pressure noise,
laser source noise (laser frequency and intensity noise),
and shot noise. 
The seismic, thermal and radiation pressure noises
can be categorized as {\it displacement noises}
since the noise sources shake up the optics directly.
These displacement noises limit the sensitivities
especially in the low-frequency regime, which
is the important region for the
ground-based gravitational wave detectors,
which target at several ten to several thousand Hz.
Recently, the idea of a gravitational wave detector
free from both the displacement noises of the optics
and the laser frequency noises at all frequencies
has been proposed in Refs.~\cite{DFI1, DFI2}.
This technique exploits the fact that the 
gravitational wave effect on the laser light takes a form different from that of optics displacements.
A signal combination
which does not sense displacement- or frequency-noises
but senses the gravitational wave contribution can be constructed
when the array of $N$ test masses consists of multiple interferometers
under the condition that $N>d+2$ ($d$ is the number of spatial
dimensions of the array).
The next paper in the series of DFI papers, Ref. \cite{DFI3},
suggested a 3-D DFI configuration consisting of four
Mach-Zehnder interferometers (MZIs).
This configuration is the most ideal
of all DFI configurations so far proposed.
%
A proof-of-principle experiment with a 2-D layout of a sub-part of
this configuration was done in Ref.~\cite{DFI4}.
There, the cancellation of displacement noise from a folding
mirror was demonstrated.
However, the 2-D layout was half that of
the 3-D DFI, and it cannot cancel
the displacements of all the optics.
In this paper, as a complete DFI system,
we report the cancellation
of the displacement both of the folding mirrors and beamsplitters
while the gravitational-wave response is retained
using the full 3-D DFI.


\begin{figure}
\begin{center}
\vspace{-2mm}
\includegraphics[width=4.5cm]{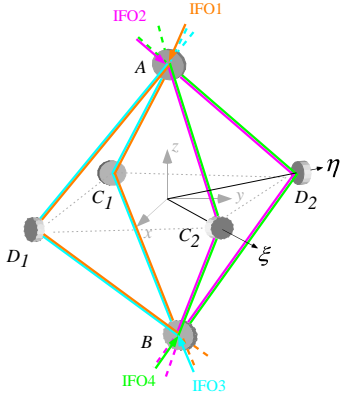}
\vspace{-1mm}
\caption{(color online). Optical topology of 3-D DFI.
The pairs of IFO$_1$ and IFO$_3$ and of IFO$_2$ and IFO$_4$
are counterpropagating to each other,
constructing two bidirectional Mach-Zehnder interferometers (MZIs).
Right and left bidirectional MZIs share the two beamsplitters.
The length from $A$ or $B$ to each folding mirror is $L$.
}
\vspace{-2mm}
\label{3DDFI}
\end{center}
\end{figure}

The 3-D DFI configuration is depicted in Fig.~\ref{3DDFI}.
Four MZIs with equal arm lengths compose the DFI.
Let us denote by IFO$_1$, IFO$_2$, IFO$_3$, and IFO$_4$
the four single MZIs drawn
in orange ($AC_1B-AD_1B$), pink ($AC_2B-AD_2B$),
light blue ($BC_1A-BD_1A$), and green ($BC_2A-BD_2A$)
in Fig. \ref{3DDFI}. Each length between a beamsplitter
and a mirror ($AC_1, AC_2, AD_1, AD_2, C_1B, C_2B,
D_1B$, and $D_2B$) is assumed as $L$
so that each interferometer is insensitive
to laser frequency noise thanks to the equal arms.
Two of the four interferometers are combined to construct
one bidirectional MZI in which two beams are counterpropagating
on the same path.
As is discussed in Ref. \cite{DFI3}, 
the DFI signal combination can be written as
\begin{eqnarray}
\label{comb}
\phi _{\rm DFI}=(\phi _1-\phi _3)-(\phi _2-\phi _4)
\end{eqnarray}
where $\phi _j (j=1,2,3,4)$ are the phase signals
of IFO$_j$.
%
%
In Eq.~(\ref{comb}), the combined signal of $(\phi _1-\phi _3)$
is the output of the bidirectional MZI
that is free from the displacements of $C_1$ and $D_1$.
Because the folding mirrors are at the midpoint
of the arms,
and the laser beams travel in both directions in the MZI,
the phase variation
due to the displacement of $C_1$ and $D_1$
arrives at each output port simultaneously.
Therefore, in the frequency domain,
IFO$_1$ and IFO$_3$ have identical
transfer functions
from the displacement of $C_1$ and $D_1$
to the signal ports, expressed as
\begin{eqnarray}
H_{1}^{C_1,D_1} (\Omega)=
H_{3}^{C_1,D_1} (\Omega)=
e^{-i L \Omega/c},
\label{eq:tf1a} 
\end{eqnarray}
where $c$ is the speed of light in vacuum,
$\omega$ is the laser frequency
and $\Omega$ the Fourier frequency of the displacement motion.
The phase modulations due to the motions of the optics
experience the same phase delay shown in Eq.~(\ref{eq:tf1a})
because the displacement sources are at the midpoint
of the path.
These identical responses to $C_1$ and $D_1$
can thus be removed by subtraction.
Similarly, $(\phi _2-\phi _4)$ is the signal of the other pair of
bidirectional MZIs, and free from the displacements
of $C_2$ and $D_2$.
Any signals that phase modulate the light and do not arise
at the exact center of the path will not arrive
simultaneously at the outputs,
and thus will not be canceled by the bidirectional MZIs.
%
%
The beamsplitter displacements are removed by
the combination of $(\phi _1-\phi _2)$ in Eq.~(\ref{comb}).
Since the laser beams of IFO$_{1}$ and IFO$_{2}$
reach the two beamsplitters simultaneously,
their responses to the beamsplitter
motion are the same.
Therefore, in the frequency domain,
IFO$_1$ and IFO$_2$ have identical transfer functions
from the displacement of $A$
to the signal ports expressed as
\begin{eqnarray}
H_{1}^{A} (\Omega)=H_{2}^{A} (\Omega)=
e^{-2i L \Omega/c}.
\end{eqnarray}
The transfer functions of the displacement motion
of beamsplitter $B$ are
$H_{1}^{B} (\Omega)=H_{2}^{B} (\Omega)=1$
since the motions of $B$ are just before the detection ports.
Therefore the beamsplitter displacements can be
eliminated by the combination of IFO$_1$ and IFO$_2$.
Similarly, the combined signal of $(-\phi _3+\phi _4)$
in Eq.~(\ref{comb}) is free from the beamsplitter displacements.
In practice, in addition to the transfer functions,
extra phase delays should be considered
that arise from the paths from the second beamsplitter
to the signal analyzer via a photo detector and signal cables.
These additional phase delays of each interferometer
should be made equal or calibrated later,
so as not to lose the balanced responses.
%

According to Ref. \cite{DFI3},
the DFI signal responds to a gravitational wave
with the transfer function,
\begin{eqnarray}
\begin{split}
H_{\rm GW}(\Omega)=
-\frac{i \omega }{\Omega}
\biggl\{ (1+1/\sqrt2)
e^{-\frac{i \Omega L}{\sqrt2 c}}
\bigl(1-e^{-\frac{2 i \Omega L(1-1/\sqrt2 )}{c}} \bigr)
\\
-(1-1/\sqrt2)
e^{\frac{i \Omega L}{\sqrt2 c}}
\bigl( 1-e^{-\frac{2 i \Omega L(1+1/\sqrt2 )}{c}} \bigr)
\biggr\}
\end{split}
\label{GWres}
\end{eqnarray}
where $\Omega$ is the Fourier frequency of the gravitational wave
and we assume the gravitational wave of $\eta-\xi$ polarization
is coming along the $z$ direction in Fig.~\ref{3DDFI}.
The 3-D DFI does not respond
to DC gravitational waves
because the response is degraded by the subtraction
when the DFI signal is built by the four MZI signals.
The DFI response is proportional to $(\Omega L/c)^2$
below the peak frequency.
The peak frequency depends on the arm length $L$,
e.g.\ when $L=1$ m, the peak is at about 150 MHz.
There have been several attempts to lower
the effective frequency of DFI:
Adding Fabry-Perot (FP) cavities
to the 3-D DFI to increase the effective path lengths
and lower the peak frequency;
another method of displacement cancellation
using one FP cavity \cite{DFIcav1, DFIcav2}.
At this moment, our $f^2$ response at low frequencies
is the best sensitivity that DFI configurations
have achieved.

\begin{figure}
\begin{center}
\vspace{-5.5mm}
\includegraphics[width=8cm]{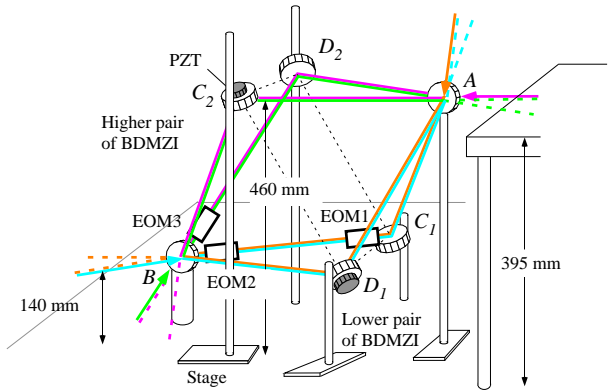}
\vspace{-3mm}
\caption{(color online). Schematic of the experiment.
The octahedron is pushed over sideways
to be set up on an optical table.
IFO$_1$ and IFO$_3$ comprise one bidirectional MZI, 
and so do IFO$_2$ and IFO$_4$.
EOM1 simulates mirror $C_1$ displacement
and EOM2 and EOM3 simulate
beamsplitter $B$ displacement.
}
\vspace{-4mm}
\label{setup}
\end{center}
\end{figure}

\begin{figure}
\begin{center}
\includegraphics[width=8cm]{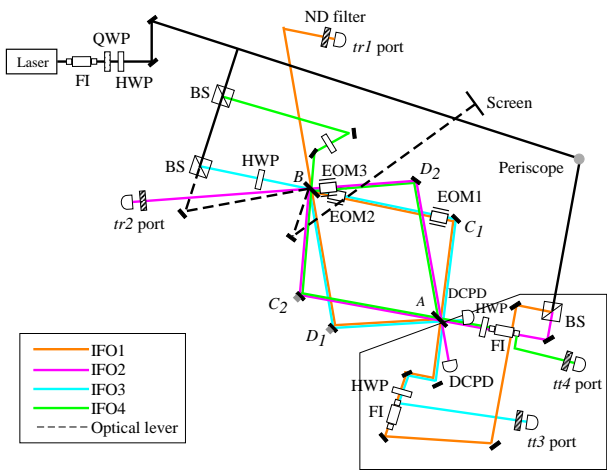}
\caption{(color online). The overhead view of the experimental setup.
HWP: half wave plate, QWP: quarter wave plate,
FI: Faraday islator, BS: beamsplitter,
DCPD: DC photo detector.
There are four interferometers:
IFO$_1$ ($A-C_1,D_1,-B$), IFO$_2$ ($A-C_2,D_2,-B$),
IFO$_3$ ($B-C_1,D_1,-C$) and IFO$_4$ ($B-C_2,D_2,-A$).
The black-dashed line indicates the optical lever method
for the alignment of 3-D configuration.
}
\vspace{-7mm}
\label{setup_2D}
\end{center}
\end{figure}


Figure.~\ref{setup}. shows
the practical setup of our experiment.
The laser source is a commercial solid-state
Nd:YAG laser at 1064 nm.
The light source is split into four by beamsplitters
so as to provide the four incident beams 
for the four MZIs.
As depicted in Fig.~\ref{setup},
IFO$_2$ and IFO$_4$ (denoted as higher interferometers)
are lying over 
IFO$_1$ and IFO$_3$ (denoted as lower interferometers).
The length $L$ was chosen to be about 40 cm
yielding peak frequency of about 370 MHz.
%
%
There are two Faraday isolators
before the lasers are sent to IFO$_1$ and IFO$_2$.
They are used for extracting the signal of IFO$_3$ and IFO$_4$
(ports $tt3$ and $tt4$~\footnote{We defined the output port
whose path experiences transmission at the first beamsplitter
and reflection at the second beamsplitter as the {\it tr}
 (transmission-reflection) port, and the other port as {\it tt}
 (transmission-transmission) port.}
%
%
%
%
exiting from the IFO$_1$ and IFO$_2$ inputs, respectively.
The signals at ports $tr1$ and $tr2$ are directly detected by the photo detectors.
The DC photo detectors (DCPD) are placed at ports $tr3$
and $tr4$ for extracting the control signals.
The extracted DC signals are sent to the servo
filters with static offsets,
then sent to the piezoelectric transducers (PZTs)
attached to mirrors $D_1$ and $C_2$.
$D_1$ and $C_2$ are controlled
by signals giving the mid-fringe locking
for the IFO$_3$ and IFO$_4$, respectively.
Once IFO$_3$ is locked, IFO$_1$ is locked automatically.
Likewise, locking IFO$_2$ automatically locks IFO$_4$
The control bandwidth of the two interferometers
is about 1 kHz.

To align the four-interferometer system,
we utilized an optical lever as shown in
Fig.~\ref{setup_2D}, using the fact that
the lower and higher sets of bidirectional MZIs
share their beamsplitters~\cite{Kthesis}.
The positions and the angles of $C_1$ and of $D_1$,
and positions and angles of $C_2$ and of $D_2$, independently
determine the position and angle of $B$
for the alignment of the higher and lower interferometer, respectively.
To make them compatible, the optical lever
was utilized to monitor the angle of beamsplitter $B$.
First, for the lower interferometer,
the angle of $B$ determined by $C_1$ and $D_1$, is adjusted
and monitored by the optical lever signal on the screen
For the higher interferometer,
we then adjust the position and angle
of $B$ determined by $C_2$ and $D_2$,
is adjusted for the lower interferometer
and displayed on the screen.
And we adjust the positions and angles of
$C_2$, $D_1$ and $A$ 
so that the two monitored marks on the screen
come to exactly the same point
which means the ideal angle of $B$.
This method achieved an almost perfect contrast.

EOMs were utilized to create phase shifts to the
laser light as simulated displacements of the optics
because the interesting frequency,
i.e., the response peak of the gravitational
wave, is expected to be in the hundred-MHz region.
To confirm the noise cancellation features of both mirror
and beamsplitter, the three EOMs were driven;
for simulating the displacement of mirror $C_1$,
EOM1 is placed at the center of the path $BC_1A$;
for simulating the displacements of beamsplitter $B$,
EOM2 and EOM3 are placed close to $B$
as beamsplitter $B$
affects both the higher and lower interferometer.
The EOMs were driven by swept-sine noise sources
provided by the RF network analyzer (Agilent 4395A).
%
The signals from the four MZIs
are detected by four fast photo detectors
(New Focus 1611) at ports $tr1, tr2, tt3$ and $tt4$.
The optical gain imbalances between the four output signals
are compensated by neutral density (ND) filters
placed in front of each detector.
The DFI signal is obtained by electrical summation
of the four signals using electric power combiners.
We adjusted the control polarity so that
$tr1$ and $tr2$ have opposite sign so as
to cancel by their summation.

\begin{figure}[t]
\begin{tabular}{c}
\begin{minipage}{1\hsize}
\begin{center}
\vspace{-3mm}
\includegraphics[width=8cm]{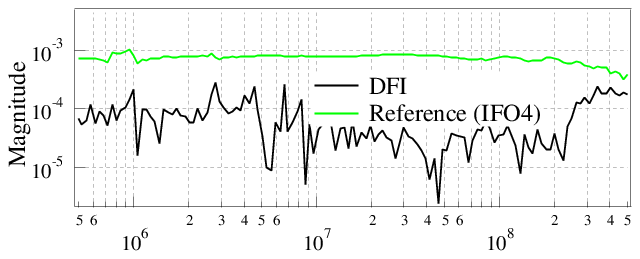}
\vspace{-12mm}
\end{center}
\end{minipage}
\\
\begin{minipage}{1\hsize}
\begin{center}
\vspace{10mm}
\includegraphics[width=8cm]{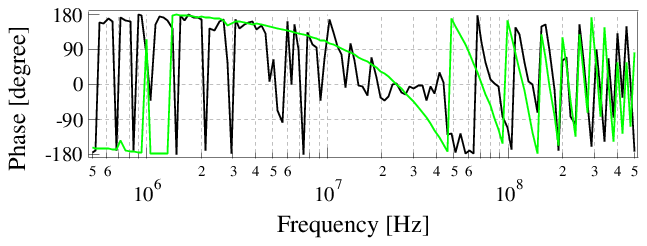}
\vspace{-3mm}
\caption{(color online).
The solid green line is the transfer function
from EOM3 to $tt4$ as a reference,
and the solid black line is the
transfer function from EOM3 to the DFI signal.
From one to two orders of magnitude of the cancellation
were attained in the DFI signal.}
\vspace{0mm}
\label{result}
\end{center}
\end{minipage}
\end{tabular}
\end{figure}

The DFI response to a gravitational wave
coming along $z$ direction was demonstrated
by multiple measurements and simulation.
Ideally, the continuous phase variation should be applied
on the entirety of the laser paths to simulate
the gravitational-wave effect.
However, only the part covered by the crystal causes
a phase shift by an EOM.
Therefore, the following procedure was adopted.
We place an EOM in the laser path
and measure the transfer function.
Then we place the EOM at the next point
and measure the transfer function, and so on.
%
The measured data are summed up
with proper phase delays that depend
on the gravitational wave direction and the EOM position.
For a gravitational wave coming along the $AB$ axis,
Eq.~(\ref{GWres}) can be approximated
by the summation of the discrete phase shifts
expressed as
%
%
\begin{align}
\begin{split}
&H_{\rm GW}(\Omega) \sim
H_{\rm sum}(\Omega)\\
&= \frac{\omega L}{cD} \biggl\{
e^{\frac{i \Omega L}{\sqrt2 c}}
\sum _{n=0}^{2D} e^{\frac{i \Omega L}{c}\frac{n}{D}}
 e^{-\frac{i \Omega L}{\sqrt2 c}\frac{n}{D}}
-e^{-\frac{i \Omega L}{\sqrt2 c}}
\sum _{n=0}^{2D}  e^{\frac{i \Omega L}{c}\frac{n}{D}}
 e^{\frac{i \Omega L}{\sqrt2 c}\frac{n}{D}}
\biggr\}
\end{split}
\label{GWsum}
\end{align}

under the condition $\Omega L/Dc\ll 1$.
$D$ is the resolution of the positions on the path
of $L$ where the phase shifts occur.
Note that the straightforward way 
to produce continuous phase shift is
to insert EOMs along all laser path.
However, that many EOMs covering laser
paths will cause a serious reduction in interferometer contrast.

\begin{figure}[t]
\begin{tabular}{c}
\begin{minipage}{1\hsize}
\begin{center}
\vspace{3mm}
\includegraphics[width=8.4cm]{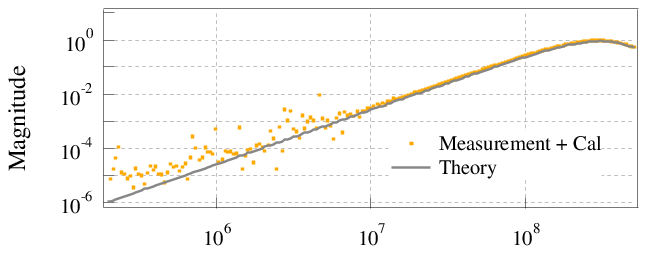}
\end{center}
\end{minipage}
\\
\begin{minipage}{1\hsize}
\begin{center}
\vspace{-3mm}
\includegraphics[width=8cm]{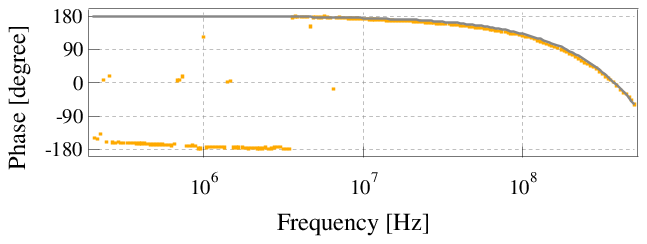}
\vspace{-3mm}
\caption{(color online). The orange dots are
the simulated DFI response to the gravitational wave
obtained by being measured and computed,
the gray solid line is the DFI response predicted
by Eq.~(\ref{GWres}). The simulated response agreed
with the theory.}
\vspace{-2mm}
\label{resultGW}
\end{center}
\end{minipage}
\end{tabular}
\end{figure}

In the experiment, 
the phase shift part of $e^{i \Omega Ln/Dc}$
in Eq.~(\ref{GWsum})
was created practically by the EOM with the discrete
resolution $D=10$.
For the first ten data, $n=0\dots 9$, 
an EOM was set in a respective position
at $nL/D$ from beamsplitter $B$, and
the respective transfer functions were  measured.
Since we cannot put an EOM on $C_1A$,
we compute the transfer function data for $n=10\dots 20$
applying proper phase delays to the data of $n=9$.
These data were summed with predicted phase delays
due to the incident angle of the gravitational wave
in accordance with Eq.~(\ref{GWsum}),
to construct the gravitational wave response.

The results for displacement-noise suppression
are shown in Fig.~\ref{result}.
The solid black line is the
transfer function from EOM1 to the DFI (combined) signal.
%
%
The solid green line is the transfer function
from EOM1 to a single MZI output (port $tt4$)
as a reference.
Comparing with these transfer functions,
we can confirm that the DFI signal has less sensitivity
to the displacement by one to two orders of magnitude.
The transfer functions in Fig.~\ref{result}. include
the response of the EOMs and the photo detectors,
and also phase delays due to the
optical and signal paths after the photo detectors to the analyzer.

%
In Fig.~\ref{resultGW}, the orange dots are the simulated
DFI response to a gravitational wave
obtained by the measurement and computing,
and the gray solid line is the DFI response predicted by
Eq.~(\ref{GWres}).
The response function of photo detectors,
and of the mechanical resonance due to the EOM crystal,
and the phase delays due to the optical paths and signal cables
from beamsplitters to the analyzer
are compensated properly in Fig.~\ref{resultGW}.
The 3-D DFI response to a gravitational wave
is well demonstrated around the peak frequency
and its $f^2$ feature can be confirmed
just below that frequency.

To summarize the results: we have successfully
constructed and operated the 3-D DFI.
The noise suppression feature of about
1 to 2 orders of magnitude
and the retained DFI response were well simulated
using EOMs over a wide frequency range.

We thank for Grant-in-Aid of the
Japan Society for the Promotion of Science
Fellows and Mitsubishi Foundation.
This research is partially supported
by the U.S. National Science Foundation.

\end{document}